# Terminal-state theorems for a two-stage Maki–Thompson rumor model: final size, stifler composition, and non-identifiability from pooled statistics

Qingquan Tang*

Wuhan Textile University, Wuhan 430200, China

*Corresponding author. E-mail: qqtang@wtu.edu.cn

**Abstract.**

Rumor models of the Daley--Kendall and Maki--Thompson family anchor the statistical mechanics of social contagion, yet the stifler composition a rumor leaves behind has resisted closed-form analysis. Inspired by the asymmetric two-step pairing (A2SP) process of step-growth polymerization, we import its common-factor cancellation into a minimal two-stage MT variant in which re-contacted spreaders are re-activated into a senior class at rate $\alpha k_1$ rather than stifled, and obtain exact terminal-state theorems. (i) Saturation: the rumor exhausts the ignorant pool for every stifling-to-recruitment ratio $\mu>0$, with a three-way tail classification and a logarithmically critical line at $\mu=1$. (ii) Final size: the terminal stifler pool splits as $(x_1,x_2)=(\mu/(\alpha+\mu),\alpha/(\alpha+\mu))$, independent of seed size; the fingerprint inverts linearly ($\alpha=\mu x_2/x_1$) and matches Gillespie simulations within 0.95%. (iii) Pooled versus history-resolved: pooled statistics are exactly $\alpha$-free on any static graph; only the history-resolved composition carries $\alpha$. Restoring the classic teller-retires channel breaks the exact closure beyond quantified gates (contact rates of 1--3% of $k_1$ for spreader channels at $\mu\leq 1.5$, up to 32% at $\mu=2$), and on configuration-model networks the composition bias decays as $\langle k\rangle$^−0.71. An exploratory inversion of 354,930 Higgs--Twitter retweet episodes, screened by an identity check and four negative controls, yields $\hat{\alpha}$ =2.21% (joint cluster-bootstrap CI [1.6%,3.0%]; ≈3.7% after calibration at the matched threshold). Terminal compositions identify this effective re-activation rate, an upper bound on contact-coupled re-activation; the mechanism itself requires timing information beyond the terminal state.

## 1. Introduction

Rumor spreading occupies a special place in the statistical mechanics of social dynamics: it is the canonical spreading process in which *contact itself can terminate transmission*. In the Daley–Kendall (DK) formulation [1], a spreader who contacts an individual that already knows the rumor loses interest; in the Maki–Thompson (MT) reformulation [2], the teller rather than the hearer retires. Both inherit the mathematical skeleton of the susceptible–infectious–removed (SIR) epidemic [3], and the analogy goes back to the first generalizations of epidemic theory to social transmission [4]. What makes the model class of practical importance today is the archival record: platform data do not merely count how far a rumor reached; they record how it ended, which accounts fell silent after a single burst, which re-activated and burst again, and when the cascade finally froze [5]. The terminal composition is thus the one quantity a platform can measure without instrumenting the process, and the one that forecasting and control actually need.

Sixty years of work have mapped this landscape qualitatively and numerically, not exactly. On the complete graph, the classical models have exact final-size and final-extent results [6–8]; on heterogeneous networks, mean-field theory yields thresholds and phase diagrams of the final fraction of ignorant [9–11] and rigorous probability supplies bounds [6]; recent reviews list a large family of variants [12]. The multi-stage reality observed on platforms (dormant accounts re-engaging, histories made of several bursts) has been modeled by enlarging the state space: the SEIZ model adds exposure and skeptic classes and is fitted numerically to Twitter cascades of news and rumors [13, 14], and forgetting-and-remembering mechanisms similarly extend the MT chain [15]. At the data end, false news is known to spread structurally farther and faster than truth [16], rumor cascades demonstrably re-ignite after dormancy [5], parameter inference for diffusion cascades is a developed toolbox [17, 18], and human communication is intrinsically bursty [19]; counter-messaging resources, in turn, are allocated through pooled, reach-type objectives [20]. Three gaps remain. (i) No closed-form history-resolved terminal law exists even at mean-field level: the final composition is known only through numerics. (ii) The multi-stage features that the data demand are exactly the features that destroy analytic tractability: one added stage destroys the closed-form history-resolved terminal composition. (iii) The identifiability question has not been asked: which observable carries which parameter, and whether the pooled statistics that intervention studies optimize can in principle determine the parameters they need. Besides, in nonlinear epidemics, aggregate data are known to identify only an equivalence class of the underlying contact structure [21], and no corresponding limit is known for rumors.

The terminal-state problem, however, was solved in a different discipline eighty years ago. For step-growth polycondensation, Flory and Carothers derived exact closed-form compositions of the reaction mixture, valid up to and including the terminal state under the equal-reactivity assumption [22, 23]. We subsequently lifted this assumption [24]: for step-growth processes with unequal reactivity, in which reaction rates depend on the history of the reacting unit, the same division still yields closed-form terminal compositions. The reason is structural, not chemical: every reaction channel of a two-step paired process (recruitment, re-activation, and stifling in the rumor system; chain growth in polycondensation) carries the same partner-pool factor. We name this mechanism asymmetric two-step pairing (A2SP): a two-stage kinetic process in which every elementary event is a pairwise contact, the two steps draw on distinct partner pools and may carry unequal reactivity, and every channel rate is multiplied by one common partner-pool factor. In rumor dynamics the common factor is the catalytic spreader pool $S = S_1 + S_2$, which multiplies every channel rate but is never itself consumed; in AA+BB polymerization it is the consumable A-monomer pool. Solvability depends only on the common factor, not on whether the common factor is consumed: writing trajectories per unit of the ignorant pool cancels the common factor and leaves linear equations, from which terminal laws and an inversion formula follow. A rumor cascade is an unequal-reactivity process in exactly this sense — re-activation is open only to spreaders with a fresh spreading history — and the common-factor cancellation transfers the same exactness to it, importing what the rumor literature lacks: closed-form trajectories, terminal-state laws, and a model-conditional estimate of the effective re-activation rate, readable from archival data.

This paper makes the minimal two-stage extension of the MT family exactly solvable and derives what exactness yields. The model differs from classical MT in one contact rule. Every spreader–spreader contact retires the contacted spreader at rate $k_2 = \mu k_1$, and a contact received by a fresh spreader $S_1$ arms it at rate $\alpha k_1$, upgrading it to the senior class $S_2$. A contacted $S_1$ thus faces the competing outcomes to re-activate or to retire, with probabilities $\alpha/(\alpha+\mu)$ and $\mu/(\alpha+\mu)$ per contact [the inflow and outflow of eq. (1b), $\gamma = \alpha + \mu$]; a contacted $S_2$ always retires. In archival terms, the two spreader classes are single-burst and multi-burst accounts. The contributions are fourfold. (a) Three terminal-state laws, with proofs: saturation with a three-way tail classification (Sec. 4.1); a *final-size law* for the history-resolved composition of the terminal stifler pool, independent of the seeding size and linearly invertible for $\alpha$ (Sec. 4.2); and a *pooled-versus-resolved dichotomy*: pooled statistics are exactly $\alpha$-free while history-resolved statistics carry a monotonic fingerprint (Sec. 4.3). (b) Two structural theorems beyond mean field: the $\alpha$-non-identifiability is architectural, holding exactly on any static contact graph by a lumpability argument (Sec. 5); and finite well-mixed populations terminate with exactly one surviving spreader frozen in the absorbing state (Sec. 6), strengthening the mean-field saturation law to an almost-sure statement. (c) The boundary of validity: we quantify how the classic stifler-contact channel (spreaders stifled by stiflers) destroys the laws at contact rates $g$ † of 1–3% of $k_1$ for the spreader-layer channels at $\mu \leq 1.5$, and up to 32% for the most robust channel (Sec. 4.3), and we map and parameterize the mean-field failure boundary on configuration-model networks [25] (Sec. 5). (d) The empirical route and its implications: an end-to-end archival inversion with a parameter-free identity check, run on 354,930 Higgs-Twitter retweet episodes (Sec. 7), and, in Sec. 8, what the theorems prescribe for measurement, calibration, and intervention in rumor research. The archival and the mechanistic routes carry the argument to the same closed form.

## 2. Model

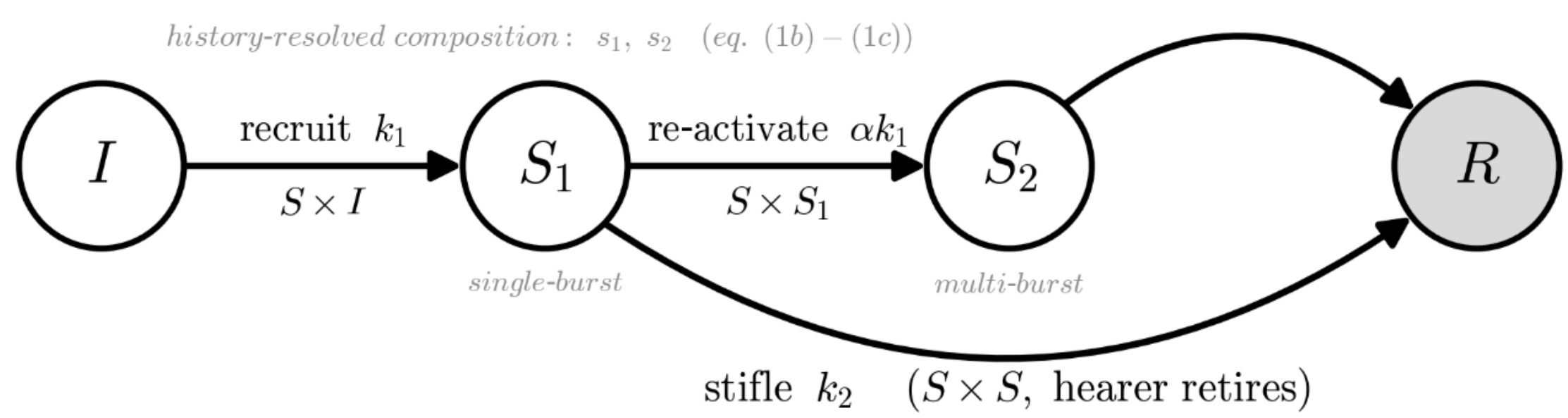

*Figure 1: The two-stage cascade. Every channel rate carries the partner-pool factor $S = S_1 + S_2$: recruitment consumes an ignorant proportional to $S$, while re-activation and stifling are spreader–spreader events: a contact received by a fresh spreader $S_1$ resolves into re-activation (rate $\alpha k_1$) or retirement (rate $k_2$) with probabilities $\frac{\alpha}{\alpha+\mu}$ and $\frac{\mu}{\alpha+\mu}$; a contact received by a senior spreader $S_2$ always retires it; the contacting spreader (the teller) always survives. The two channel rates $\alpha k_1$ and $k_2 = \mu k_1$ sum to the packaged fresh-spreader rate $k_1\gamma = \alpha k_1 + k_2$ of eq. (1b). R is absorbing. The model is population-normalized, $N = 1$; dimensionless groups are $\mu = k_2/k_1$ and $\gamma = \alpha + \mu$. Archival reading: $S_1$ denotes single-burst accounts and $S_2$ re-activated, multi-burst accounts. The shaded band marks the history-resolved composition ($s_1, s_2$; eq. (1b)–(1c)); the ruled baseline beneath marks the pooled layer ($i, S, r$; eq. (1a),(1d)–(1e)). The anatomy of a single episode is given in Fig. 2(a).*

Individuals move through the chain $I \to S_1 \to S_2 \to R$ (Fig. 1). A spreader of either class ($S_1$ or $S_2$) who contacts an ignorant recruit it at rate $k_1$ per contact; the recruit enters as a fresh $S_1$. A spreader who is contacted by another spreader is *stifled* (it retires to the absorbing class $R$) at rate $k_2$; the teller survives (catalytic contact). The two-stage variant of classical MT is the *re-activation* channel: a fresh spreader $S_1$ who is contacted by any spreader is not stifled but promoted to $S_2$ at rate $\alpha k_1$. The re-activation rate $\alpha$ measures how strongly re-exposure promotes a recruit: how far a second contact re-engages an account that has forwarded once. In archival terms, $S_1$ are accounts that have forwarded once and not yet re-engaged; $S_2$ are re-activated, multi-burst accounts; $R$ are silenced accounts.

Writing population fractions $(i, s_1, s_2, r)$ with $i + s_1 + s_2 + r = 1$ and $S \equiv s_1 + s_2$, the mean-field equations are

$$\dot{i} = -k_1 i S, \qquad (1a)$$

$$\dot{s}_1 = k_1 S i - k_1 \gamma S s_1, \qquad \gamma = \alpha + \mu, \qquad (1b)$$

$$\dot{s}_2 = \alpha k_1 S s_1 - k_2 S s_2, \qquad (1c)$$

$$\dot{S} = k_1 S i - k_2 S^2, \qquad (1d)$$

$$\dot{r} = k_2 S^2. \qquad (1e)$$

The fresh-spreader equation (1b) packages the two spreader–spreader outcomes of a fresh spreader into a single clock: a contact received by $S_1$ resolves at rate $k_1$, into re-activation with probability α/γ or stifling with probability μ/γ (γ=α+μ); the packaged rate $k_1\gamma$ is the sum of the two channel rates, αk1 and $k_2=\mu\, k_1$, as drawn in Fig. 1.

Two structural facts follow immediately and organize what follows. First, the merged pool layer (1d)–(1a) is a closed two-dimensional system that does not contain $\alpha$ at all: the rumor-size dynamics is $\alpha$-free for every value of the re-activation rate. Second, the reach dynamics (1e) depends only on $S$, so the cumulative reach $r(t)$ is likewise $\alpha$-free. All $\alpha$-dependence is confined to the *history-resolved layer* (1b)–(1c). At $\alpha = 0$ the re-activation channel closes ($S_2$ never populates) and the pooled layer reduces to the classical MT/DK mean field; $\alpha = 1$ equalizes the re-activation and recruitment rates. In this sense, the pooled layer is MT/DK-equivalent for all $\alpha$, and $\alpha$ reshapes only the history-resolved composition, a point that becomes a theorem in Sec. 4.3 and 5.

The two-stage variant is not an arbitrary extension. (i) Its α=0 core is the classical Maki–Thompson chain: the epidemic-theoretic modeling of idea transmission by Goffman and Newill [4] and the Kermack–McKendrick threshold framework [3] precede it; the Daley–Kendall stochastic process [1] and its textbook formulation [2] define it; and Sudbury's exact final size [6] together with the network generalizations of Moreno et al. [9] and Nekovee et al. [10] solve its terminal state — every classical exact result is therefore recovered as a special case. (ii) The re-activation channel is the model's single innovation, and it encodes what archival studies repeatedly document: individual-level bursty re-activation and successive bursts of rumor popularity [19, 25], with rumor cascades measurable in depth, size, and structural virality [5, 16]; memory-type mechanisms inside the spreader class — forgetting and remembering — were introduced by Zhao et al. [15], and the senior class $S_2$ is exactly such a remembered state, entered by re-exposure at rate $\alpha k_1$. (iii) The payoff of the split is methodological: the pooled layer (1a), (1d)–(1e) closes without $\alpha$, the history-resolved layer (1b)–(1c) carries $\alpha$ alone, and this separation is what makes the terminal-state theorems of Secs. 4–5 and the archival inversion $\alpha=\mu x_2/x_1$ possible — a competing-risks bookkeeping (re-engagement versus silencing after re-exposure) that classical MT does not track.

## 3. Closed-form trajectories by common-factor cancellation

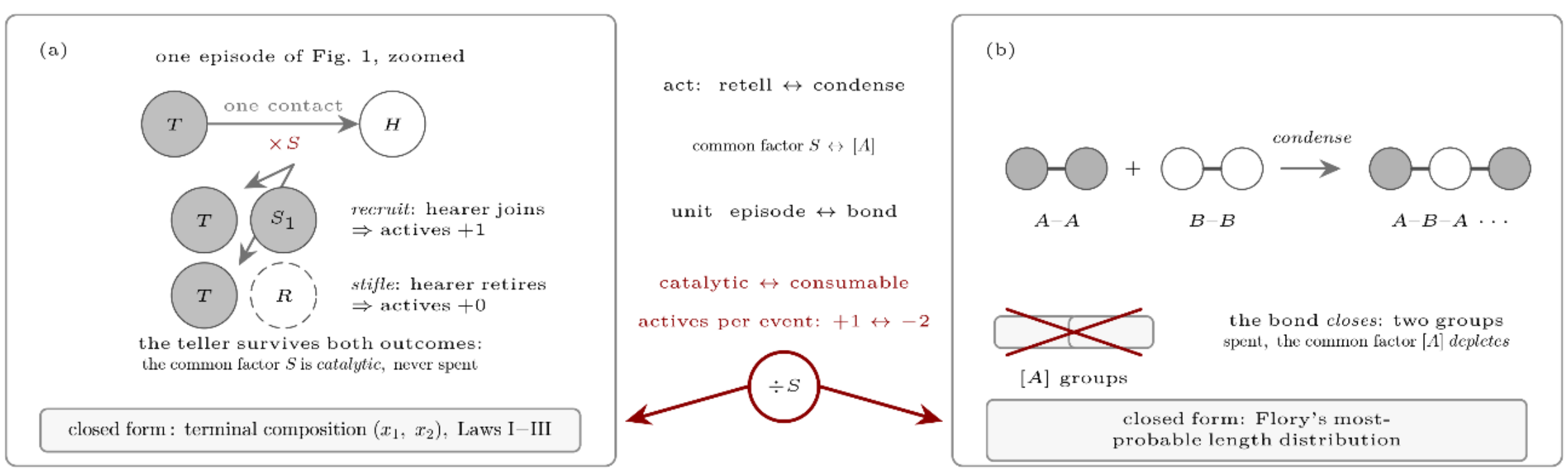

*Figure 2: The polycondensation correspondence. (a) One episode of the Fig. 1 cascade: the teller survives both outcomes, so the common factor S is catalytic. (b) AA+BB step-growth polycondensation (Flory–Carothers, 1936): each bond spends two functional groups, so the common factor [A] depletes. Dividing every rate by S (÷S) absorbs the one structural difference and transfers Flory's exactness to the terminal composition (Sec. 4).*

The division performed below is the rumor-side instance of the kinetic logic of step-growth polycondensation [21–23], and it is the foundation of every closed form obtained in this paper. In the AA+BB polycondensation treated by Flory and Carothers, all reaction channels carry one and the same partner-pool factor, the A-group balance [A], and dividing the rate equations by that balance converts the nonlinear kinetics into linear composition equations, exact up to and including the terminal state under equal reactivity [21, 22]; our unequal-reactivity extension [23], in which rates depend on the history of the reacting unit, keeps them in closed form, precisely the form the rumor setting requires. The rumor cascade shares this skeleton (Fig. 2): every channel (recruitment, re-activation, stifling), carries the same factor S, with one structural difference: in polycondensation the common factor is consumed; here it is catalytic. The difference is absorbed by the division itself: trajectories per unit of the ignorant pool, obtained by dividing by (1a), cancel S and leave the linear system (2). What Flory obtained for chain-length distributions, the same move delivers here: the closed-form trajectories (3), the exact terminal history-resolved composition (Laws I–III), and the inversion $\alpha=\mu x_2/x_1$, each a corollary of the common-factor cancellation, and each exact for every α and μ [23].

Divide (1d) by (1a) and (1b) by (1a): the partner-pool factor $S$, the common factor, cancels completely (Fig. 2),

$$\frac{dS}{di} = -1 + \mu\,\frac{S}{i}, \qquad \frac{ds_1}{di} = -1 + \gamma\,\frac{s_1}{i}, \qquad (2)$$

leaving two linear inhomogeneous equations driven only by the dimensionless groups $\mu$ and $\gamma = \alpha + \mu$. With initial condition $(i_0, s_1, s_2) = (1-\varepsilon, \varepsilon, 0)$ at $t = 0$ and $k_1$ the rate unit, the solutions for $\mu, \gamma \neq 1$ are

$$S(i) = C_S\, i^{\mu} - \frac{i}{1-\mu}, \qquad C_S = \frac{\varepsilon + i_0/(1-\mu)}{i_0^{\mu}}, \qquad (3)$$

$$s_1(i) = C_1\, i^{\gamma} - \frac{i}{1-\gamma}, \qquad C_1 = \frac{\varepsilon + i_0/(1-\gamma)}{i_0^{\gamma}}, \qquad (4)$$

with logarithmic branches $S = i(C_S - \ln i)$ $[C_S = \varepsilon/i_0 + \ln i_0]$ and $s_1 = i(C_1 - \ln i)$ at $\mu = 1$ and $\gamma = 1$. Both closed forms are exact for every parameter combination; the logarithmic branches are the only exception, confined to the resonant lines (Sec. 4.1).

The pooled-versus-resolved statistics are read off the history-resolved layer. Define $x_1$ ($x_2$) as the fraction of the population that was stifled *as* a fresh (re-activated) spreader, $x_k = \int k_2 s_k S\, dt$. These are the archival observables: $x_1$ counts accounts silenced after a single burst, $x_2$ counts accounts silenced after re-activation. Because the stifling flux divides by the same common factor as the ignorant flux, the resolved composition accumulates along the trajectory as

$$x_k = \mu \int_0^{i_0} \frac{s_k(i)}{i}\, di, \qquad (5)$$

a second, independent common-factor cancellation. Equations (4) and (5) determine the terminal history-resolved composition exactly.

## 4. Terminal-state laws

### 4.1 Law I: saturation and the tail trichotomy

We begin with the fate of the rumor itself. The mean field (1) saturates for every parameter choice, and the approach to the frozen state falls into three regimes.

**Theorem 1.** For all $\mu > 0$ the mean field (1) satisfies $(i, s_1, s_2) \to (0,0,0)$ and $r \to 1$ as $t \to \infty$, and the approach is asymptotic (i.e., $i(t) > 0$ for all finite $t$). Moreover, with $C_S$ from (3),

$$i(t) \sim \begin{cases} (\mu k_1 C_S t)^{-1/\mu}, \quad S(t) \sim (\mu k_1 t)^{-1}, \quad S/i \to \infty, & \mu < 1, \\ (k_1 t \ln t)^{-1}, \quad S/i = C_S + \ln(1/i) \to \infty, & \mu = 1, \\ (\mu - 1)/(k_1 t), \quad S(t) \sim (k_1 t)^{-1}, \quad S/i \to 1/(\mu - 1), & \mu > 1. \end{cases} \qquad (6)$$

The proof (Appendix A) combines the closed form (3) with an exact identity, $d(i^{-\mu})/dt = \mu k_1 \left(C_S - i^{1-\mu}/(1-\mu)\right)$, whose right-hand side converges to the positive constant $\mu k_1 C_S$. We emphasize two points. First, saturation holds for *all* $\mu > 0$: the closed form (3) keeps $S(i)$ positive on the whole interval $(0, i_0]$ for every $\mu$ (the ratio $S/i$ has exactly the trichotomy limits of (6)), so the recruitment flux $k_1 \cdot i \cdot S$ vanishes only as $i$ itself does, and the induced decay $di/dt = -k_1 \cdot i \cdot S(i)$ is a power law (logarithmic at $\mu = 1$) whose zero is approached only as $t \to \infty$; the frozen condition $i = 0$ of the trajectory framework is therefore structurally guaranteed. Second, the tail classification is *not* a finite-time-versus-asymptotic dichotomy, since saturation is asymptotic in continuous time for every $\mu$; it is a genuine three-way classification of *who outlives whom*: for $\mu < 1$ the ending is spreader-led ($S/i \to \infty$: the pool evaporates slowly while ignorance collapses faster, $t^{-1/\mu}$ with exponent $> 1$); for $\mu > 1$ the two decay together with the universal ratio $1/(\mu - 1)$; and $\mu = 1$ is logarithmically critical, $i \sim 1/(k_1 t \cdot lnt)$, separating the two regimes rather than belonging to either. Numerically, the exact identities behind (6) hold to $\leq 6 \times 10^{-10}$ along integrated trajectories, and the linearized tail slopes match $+1.000000$ to $1.4 \times 10^{-5}$ (Fig. 3); this also settles, against an earlier heuristic, that no finite-time freezing occurs for $\mu < 1$.

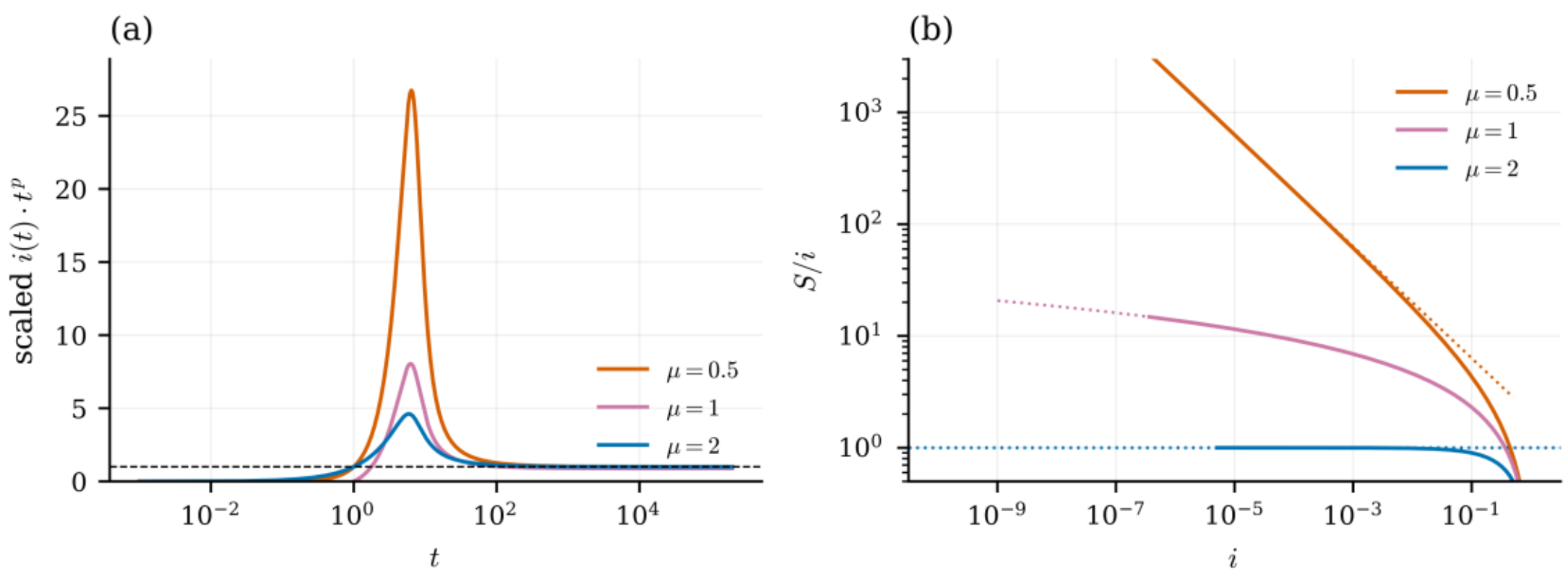

*Figure 3: Law I. (a) Tail collapse:* $i(t)$ *scaled by the predicted tail factors of Eq. (6) for* $\mu = 0.5$ *(*$i\,(\mu k_1 C_S t)^{1/\mu}$*),* $\mu = 1$ *(*$i\,k_1 tlnt$*),* $\mu = 2$ *(*$i\,k_1 t$*); plateaus at 1 confirm the trichotomy. (b) Phase-plane asymptotics:* $S/i$ *against* $i$*; the spreader-led branch (*$\mu < 1$*) diverges as* $C_S\, i^{\mu - 1}$*,* $\mu = 1$ *grows logarithmically, and the co-decay branch (*$\mu > 1$*) settles at the constant* $1/(\mu - 1)$ *(dotted guides).*

### 4.2 Law II: the final-size law

The second law fixes what the rumor leaves behind. The terminal stifler pool splits by spreading history, and the split forgets the seeding size entirely.

**Theorem 2.** For all $\mu, \alpha > 0$ and all seeding sizes $0 < \varepsilon < 1$, the terminal history-resolved composition of the mean field (1) is

$$x_1 = \frac{\mu}{\alpha + \mu}, \qquad x_2 = \frac{\alpha}{\alpha + \mu}, \qquad x_1 + x_2 = 1. \qquad (7)$$

By (5) and the trajectory equation (2), $s_1/i = (1 + ds_1/di)/\gamma$; since $s_1(i) = C_1 i^\gamma - i/(1-\gamma)$ is absolutely continuous on $[0, i_0]$ with integrable derivative and $s_1(0^+) = 0$, $s_1(i_0) = \varepsilon$,

$$x_1 = \frac{\mu}{\gamma}\int_0^{i_0}\left(1 + \frac{ds_1}{di}\right)di = \frac{\mu}{\gamma}[i_0 + \varepsilon - s_1(0^+)] = \frac{\mu}{\gamma},$$

using $\varepsilon + i_0 = 1$. The merged version of the same argument gives $x_1 + x_2 = \mu \int_0^{i_0}(S/i)\, di = (i_0 + \varepsilon) = 1$; hence $x_2 = 1 - x_1 = \alpha/\gamma$.

The proof is uniform in the parameters (no branch split at $\gamma = 1$), and the seeding size $\varepsilon$ cancels exactly, not perturbatively: the rumor "forgets its seeding and remembers only its dynamics." Two corollaries make the law operational:

**Corollary 3.** $\alpha = \mu\, x_2/x_1$: the re-activation rate is recoverable from the terminal history-resolved composition alone, by a linear estimator.

**Corollary 4.** The terminal composition depends on $(\alpha, \mu)$ only through the ratio $\alpha/\mu$ ($(x_1, x_2) = (1, \alpha/\mu)/(1 + \alpha/\mu)$), whereas the history-resolved trajectory depends on them through the sum $\gamma = \alpha + \mu$ and the merged trajectory only through $\mu$. The three layers of the solution "see" the two parameters through different lenses: ratio, sum, and one knob, respectively. In particular, iterated limits need not commute with joint ones: along paths $\alpha/\mu \to c$ the composition converges to $\frac{c}{1+c}$ for any $c$.

Gillespie simulation [27] ($N = 6 \times 10^4$, direct method, six seeds per point) confirms (7) over the grid $(\alpha, \mu) \in \{0.5, 1, 2\}^2$ with a maximum deviation of 0.95% (inside the $\pm 2\%$ acceptance gate set before the runs), and a seed-size scan (seed count $\varepsilon N = 60 \to 6000$ at fixed $N$) shows the $x_1$ deviations shrinking monotonically, 0.95% → 0.28%, verifying the seeding independence (Fig. 4).

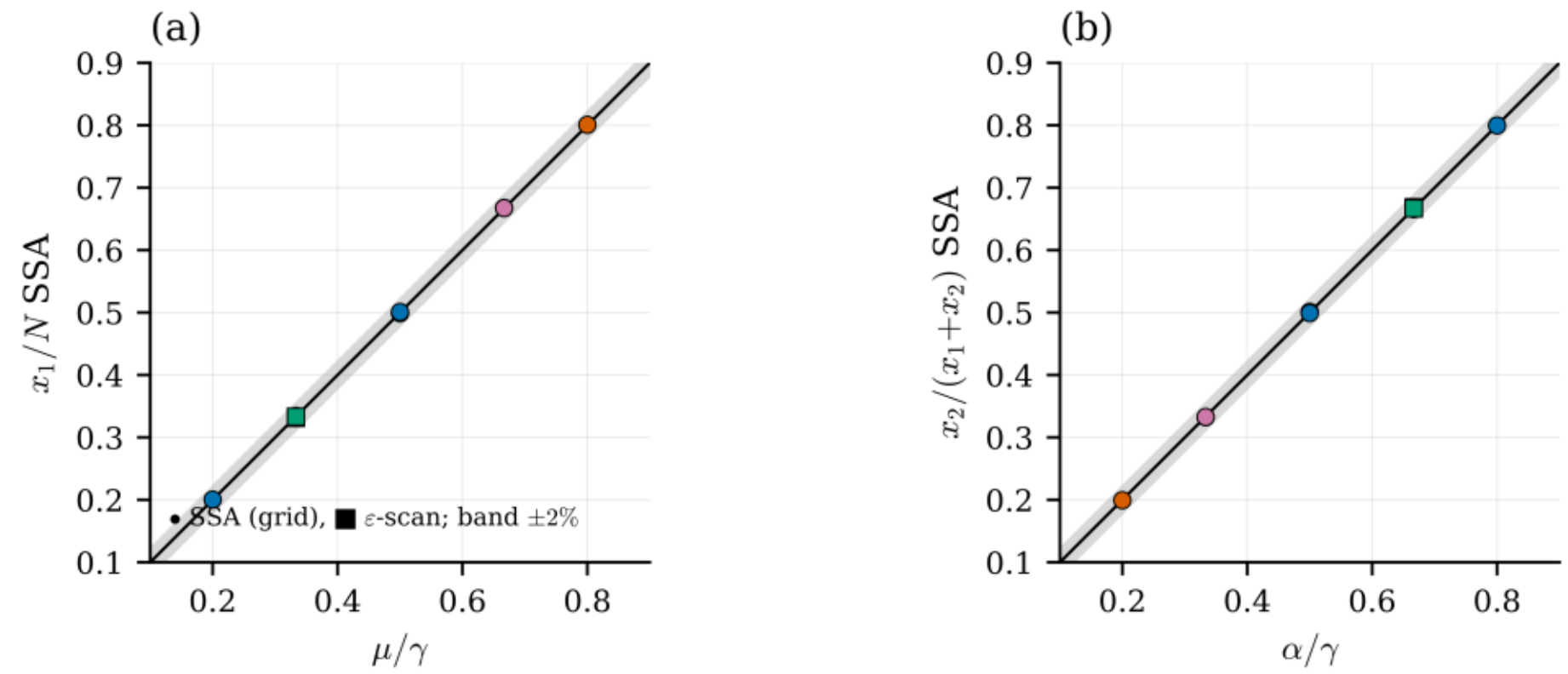

*Figure 4: Law II. Fingerprint calibration: simulated terminal history-resolved composition against the final-size law (7). (a) Left: the history-resolved composition $x_1/N$ versus $\mu/(\alpha+\mu)$; (b) right: the stifler partition share $x_2/(x_1+x_2)$ versus $\alpha/(\alpha+\mu)$. All nine parameter points (circles) and three seeding sizes (squares) fall within the $\pm 2\%$ band.*

### 4.3 Law III: pooled versus history-resolved statistics, and the price of the classic channel

The third law separates two records of the same run: the pooled trajectory, identical for every $\alpha$, and the history-resolved composition, which remembers $\alpha$ alone.

**Theorem 5.** (i) The merged trajectory $S(t)$ (hence $i(t)$ and $r(t)$) is exactly independent of $\alpha$. (ii) The history-resolved trajectory (4) depends on $\alpha$ only through the sum $\gamma = \alpha + \mu$, and the terminal composition only through the ratio $\alpha/\mu$. (iii) $\alpha$ is identifiable from the resolved composition (Cor. 3) and from no pooled statistic.

Part (i) is immediate from (1d); parts (ii)–(iii) from (4), (7). The dichotomy, in which the dynamics forgets $\alpha$ while only the resolved composition remembers it, is what makes the archival inversion of Cor. 3 both possible and non-redundant: time series of reach carry no information about re-activation, while the terminal history-resolved composition carries it all.

How much of this survives the classic MT stifling convention? In MT, a spreader is also stifled upon contacting a *stifler*. Adding this channel at strength $g$ (per $S \times R$ contact, teller retires) destroys the common-factor cancellation: the history-resolved layer acquires the term $g\, s_1 r/(iS)$ and becomes three-dimensional. Nevertheless the merged layer survives in closed form. Substituting the conservation $r = 1 - i - S$ into $\dot{S} = k_1 S i - k_2 S^2 - g k_1 S r$ gives

$$\frac{dS}{di} = -(1+g) + (\mu - g) \cdot \frac{S}{i} + \frac{g}{i}, \quad (8)$$

a two-dimensional *linear* equation in $S(i)$ whose integration constant C is fixed by $S(i_0)=\varepsilon$, with solution

$$S(i; g) = C \cdot i^{\mu - g} + \frac{g}{g - \mu} - \frac{1+g}{1+g-\mu} \cdot i, \quad (9)$$

which reduces to (3) as $g \to 0$ (residual 0 by computer algebra; the resonant lines $g = \mu$ and $g = \mu - 1$ carry logarithmic branches, as at $\mu = 1$). Three consequences follow. (i) *The merged layer stays exactly $\alpha$-free for every $g$*: Eq. (8) contains no $\alpha$; numerically, trajectories at $\alpha = 0.5$ and $\alpha = 3$ differ by $\leq 3 \times 10^{-12}$ at all $g$ tested. The excluded channel does not erode the non-identifiability; it erodes saturation and the resolved composition. (ii) *Saturation breaks*: $S(i; g)$ now crosses zero at a positive $i_f(g)$ (the ignorant pool freezes partially, the classic MT behavior), with $i_f \sim g^{1/\mu}$ scaling for $\mu < 1$ and $i_f \sim (\mu - 1)g/\mu$ for $\mu > 1$; Eq. (9) reproduces the numerically integrated $i_f(g)$ to six decimals. (iii) *The resolved composition degrades fast*: the history-resolved composition departs from (7) by more than 2% once $g \gtrsim g^\dagger(\mu)$: the stifler share at $g^\dagger \approx 0.007$–$0.029$ and the spreading history $x_1$ at $g^\dagger \approx 0.015$–$0.029$ for $\mu \leq 1.5$, with the $x_1$ gate rising to 0.317 at $\mu = 2$ (exact mean-field values in Table 1 in Appendix C). The classic channel is therefore not a benign perturbation of the two-stage model, and the exclusion required for the closed forms is quantitatively justified.

## 5. Networks: architectural non-identifiability and the failure boundary

### 5.1 Non-identifiability is architectural

In the mean field the pooled layer sheds $\alpha$ by explicit cancellation. On a contact graph no such cancellation is available, yet the non-identifiability survives for a structural reason.

**Theorem 6.** On any fixed contact graph, the continuous-time Markov chain of Sec. 2 (recruitment on $S$–$I$ edges at rate $k_1$; stifling on ordered spreader–spreader pairs at rate $k_2$; re-activation on ordered $S \to S_1$ pairs at rate $\alpha k_1$) has a pooled process $(i, S, r)$ and, more strongly, every *pooled edge-type* count, whose generator does not depend on $\alpha$. Here a pooled edge-type count tallies edge occurrences aggregated over the $S_1/S_2$ labels: $I$–$S$, $S-S$, $S-R$, $I-I$, $I-R$ (and their time-integrals). Consequently all pooled statistics, including the full trajectories $i(t)$, $S(t)$, $r(t)$ and the terminal $(i_f, n_f)$, are exactly $\alpha$-free. Label-resolved edge counts (e.g. $S_1-S_2$ contacts) are not pooled and do carry $\alpha$, consistently with the history-resolved statistics of Sec. 4.3. The same holds for time-dependent graphs whose dynamics is independent of the $S_1/S_2$ labels, and class-wise for the heterogeneous mean-field system.

The proof (Appendix B) is a lumpability argument: re-activation moves an agent between $S_1$ and $S_2$ without touching the pool or any edge type, while recruitment and stifling rates are symmetric within the pool. The non-identifiability that held by cancellation in mean field is thus *architectural*: it is a property of the channel bookkeeping, not of homogeneous mixing. Numerically, on a Poisson configuration model ($\langle k \rangle = 8$, $N \approx 2 \times 10^4$, $\mu = 0.5$, four graphs with four SSA seeds each), the mean pooled trajectories at $\alpha = 0.5$ and $\alpha = 3$ overlap within Monte Carlo error (CI overlap 100%; terminal $i_f/N$: 0.0246 vs. 0.0248) while the terminal history-resolved composition changes markedly, $x_2/(x_1 + x_2)$: 0.437 at $\alpha = 0.5$ and 0.827 at $\alpha = 3$ (Fig. 5).

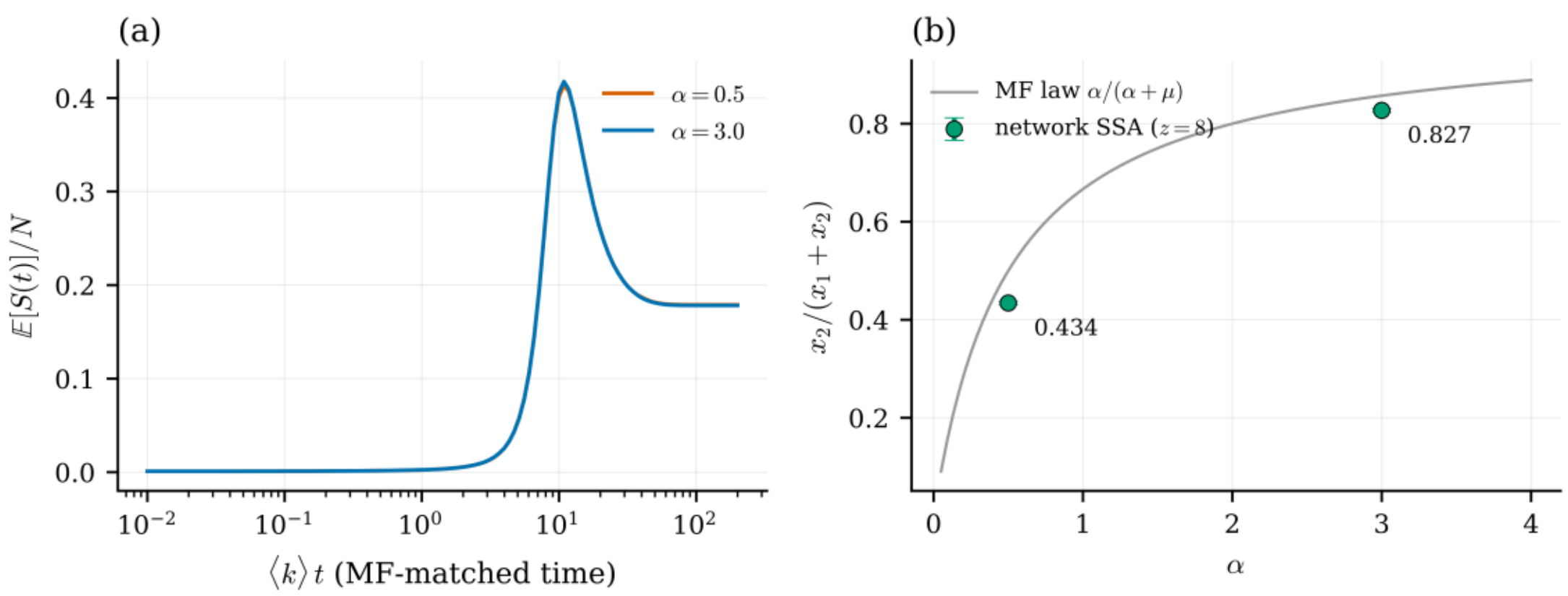

*Figure 5: Pooled versus history-resolved statistics on networks (Poisson* $\langle k \rangle = 8$*,* $\mu = 0.5$*). (a) Mean pooled trajectories* $E[S(t)]$ *for* $\alpha = 0.5$ *and* $\alpha = 3$ *on the same random graph: indistinguishable within confidence bands (CI overlap* 100%*), as Theorem 6 requires. (b) The terminal history-resolved composition along the same runs: the history-resolved composition shifts with* $\alpha$ *(MF law, gray).*

### 5.2 Terminal structure and the failure boundary

On networks the frozen state is no longer empty, and the exact laws no longer hold exactly. The proposition below identifies what survives; the simulations that follow measure how much bends.

**Proposition 7.** On a static graph, an absorbing state of the pooled dynamics is reached exactly when (i) no $S-I$ edge remains and (ii) the remaining spreaders form an independent set of the graph. Consequently, the terminal state carries a two-dimensional residual $(i_f, n_f)$, unreached ignorants *and* isolated spreaders, both absent in mean field, where $(i_f, n_f) = (0,0)$. Here n_f counts surviving spreaders (a count, not a fraction); an isolated spreader has no spreader–spreader contact and can neither stifle, be stifled, nor be re-activated.

Mechanically, the rumor freezes when the last ignorant are shielded (all their spreading neighbors have retired) and the surviving spreaders no longer touch each other. Simulations on Poisson configuration models ($N = 2 \times 10^4$, eight seeds per point, $\alpha = 1$) map the failure boundary (Fig. 6): the reach $R_f/N$ climbs from 0.65 at $\langle k \rangle = 4$ to 0.95 at $\langle k \rangle = 40$, still short of the mean-field value 1; the frozen fractions decay with $\langle k \rangle$; and the resolved composition deviation from the final-size law closes as a power law, $\propto \langle k \rangle^{-0.71}$ (bootstrap 95% CI $[-0.81, -0.54]$ over graph realizations; $n = 10$ realizations $\times$ 3 runs per point, Fig. 7). Heterogeneous mean field (HMF) computed on the same degree distributions is nearly saturated ($R_f/N \geq 0.98$ everywhere) and *over*estimates the history-resolved share in the opposite direction ($+38\%$ at $\langle k \rangle = 6$): the annealed approximation misses both the shielding and the sign of the resolved composition shift. Degree heterogeneity at fixed mean degree makes things worse, not better: at $\langle k \rangle = 7.4$ and $7.1$, scale-free graphs ($\tau = 2.5, 3$) deviate further from the final-size law than Poisson at the same mean degree ($-13.2\%$ and $-11.4\%$ vs. $-8.9\%$). The direction is the opposite of SIR intuition, and the mechanism is bookkeeping: stifling is a spreader–spreader event, whose total rate scales with $\langle k^2 \rangle$ and is fed faster than recruitment by heavy tails. The analytic root of all mean-field failure is visible in the proof of Theorem 2: the common-factor-cancellation identity requires a single driver variable, and degree heterogeneity splits the driver into classes. The deviation is systematic in the parameters (Fig. 7b): it grows with $\mu$ ($8.5\% \to 13.7\%$ at $\langle k \rangle = 8$ for $\mu = 0.5 \to 2$; stifling is the $\langle k^2 \rangle$-weighted event), decreases with $\alpha$ ($13.3\% \to 3.4\%$ for $\alpha = 0.5 \to 3$), and depends only weakly on $N$ ($8.5\% \to 7.9\%$ for $N = 2 \times 10^4 \to 5 \times 10^3$).

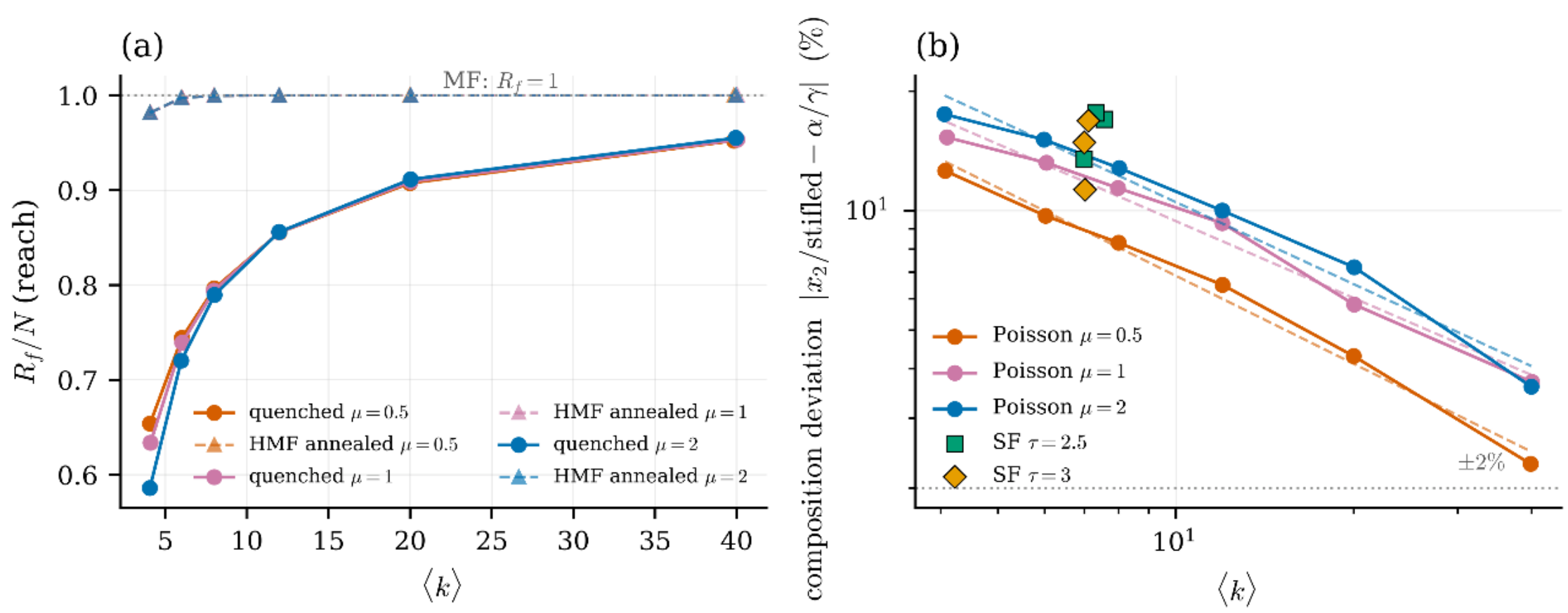

*Figure 6: Mean-field failure boundary on configuration models ($N = 2 \times 10^4$, $\alpha = 1$; eight seeds per point). (a) Terminal reach $R_f/N$ versus $\langle k \rangle$: quenched simulation (circles) against HMF (triangles, near-saturated) and the MF value* 1. *(b) Composition deviation*

$|x_2/(x_1+x_2)-\alpha/\gamma|$ *versus* $\langle k\rangle$ *(log–log): Poisson points close as* $\langle k\rangle^{-0.71}$ *(dashed guides); scale-free graphs at matched* $\langle k\rangle$ *(diamonds) deviate more.*

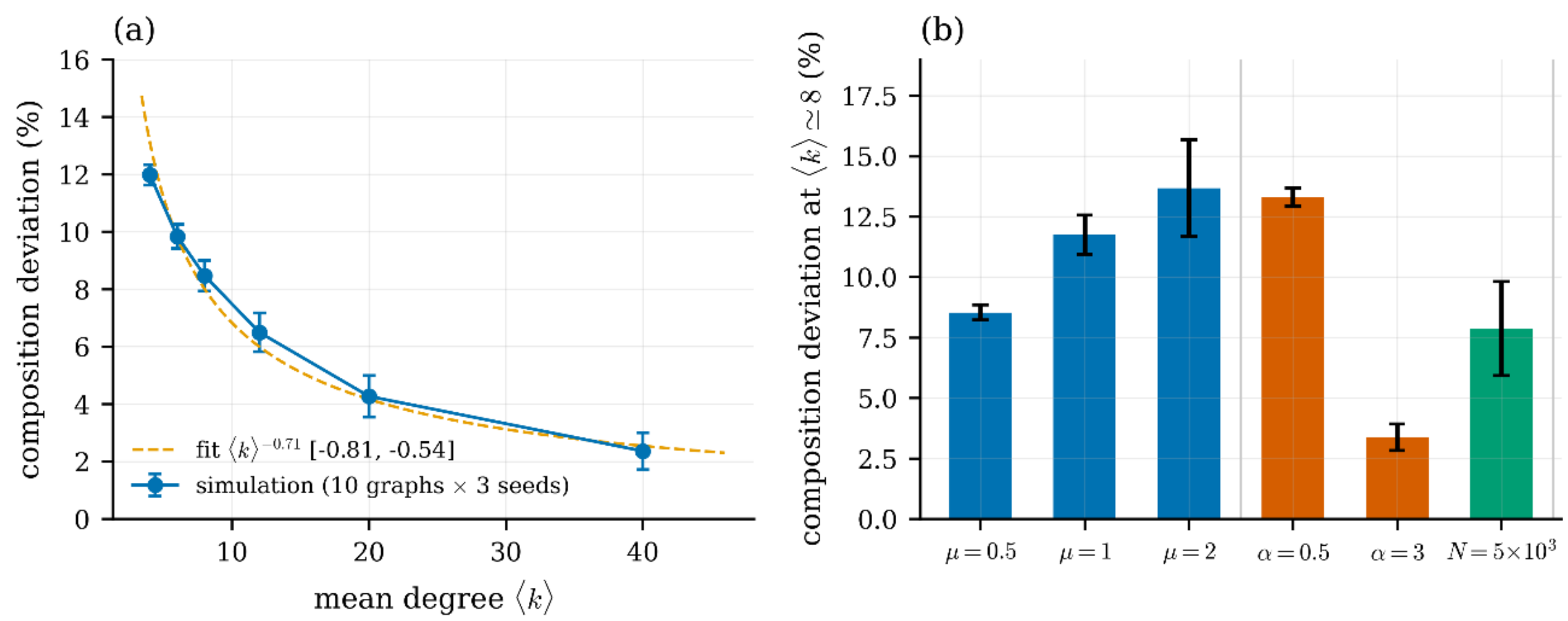

*Figure 7: Composition deviation under graph uncertainty and parameter sweeps. (a) Failure boundary with confidence intervals: composition deviation* $|x_2/(x_1+x_2)-\alpha/(\alpha+\mu)|$ *versus mean degree* $\langle k\rangle$ *(Poisson configuration model,* $N = 2\times 10^4$, $\mu = 0.5$, $\alpha = 1$*); each point averages 10 independent graph realizations × 3 seeds, with ± bootstrap* 95% *CIs; the dashed curve is the power-law fit, exponent* $-0.71$ *(*95% *CI* $[-0.81, -0.54]$*). (b) Parameter dependence at* $\langle k\rangle \simeq 8$*: the deviation grows with* $\mu$ *(blue,* $\alpha = 1$*), decreases with* $\alpha$ *(orange,* $\mu = 0.5$*), and depends only weakly on* $N$ *(green,* $\mu = 0.5$, $\alpha = 1$*).*

## 6. Finite *N*: the lone spreader

In a finite population the process terminates in finite time, and the way it terminates sharpens Law I rather than spoiling it.

**Theorem 8.** In the well-mixed stochastic process with $N \geq 2$, any initial seed count $n_0 \geq 1$, and any $\mu > 0$, the terminal state satisfies

$$\mathbb{P}(i_f = 0) = 1, \qquad \mathbb{P}(S_f = 1) = 1,$$

and consequently $R_f = \frac{N-1}{N}$ exactly: the final-size identity $x_1 + x_2 = 1$ holds up to the exact finite-population correction $x_1 + x_2 = 1 - \frac{1}{N}$.

The stifling event fires on each ordered spreader–spreader contact, at total rate $k_1 \cdot S(S-1)/N$; it carries the factor $S-1$: a lone spreader has no pairing partner and cannot retire, while its recruitment rate $k_1 i\, S_{count}/N$ is positive whenever $i > 0$. Hence (i) $i_f > 0$ is impossible, because recruitment persists at positive rate until the ignorant pool is exactly empty; and (ii) $n_f = 0$ is impossible, because the pool count decrements by one per stifling but never from one to zero. Absorption occurs only at $(i, n) = (0, 1)$, which is reached a.s. because $i$ decreases along a finite chain of recruitment events.

The theorem strengthens Law I beyond mean field: for the *driver* variable, the stochastic system saturates *exactly*, not merely asymptotically. The true finite-$N$ defect is not a nonzero probability of failing to saturate but a single isolated spreader that, with no partner to stifle it and no ignorant left to recruit, persists in the absorbing state: the process has terminated, and the survivor can neither recruit nor be stifled further. The $-1/N$ correction to the resolved composition is the fingerprint's finite-population bias. Simulation across 36 configurations ($N \in \{10^2, \ldots, 3 \times 10^4\}$, $\mu \in [0.5, 3]$, up to 4000 replicates each) confirms both statements without exception, and measures the correction against its exact prediction $1/N$ (Fig. 8). On networks the proof's premise fails (recruitment *can* be structurally blocked), and there the residual $i_f > 0$ of Proposition 7 appears; the dividing line between the two worlds is whether the recruitment channel can be topologically frozen.

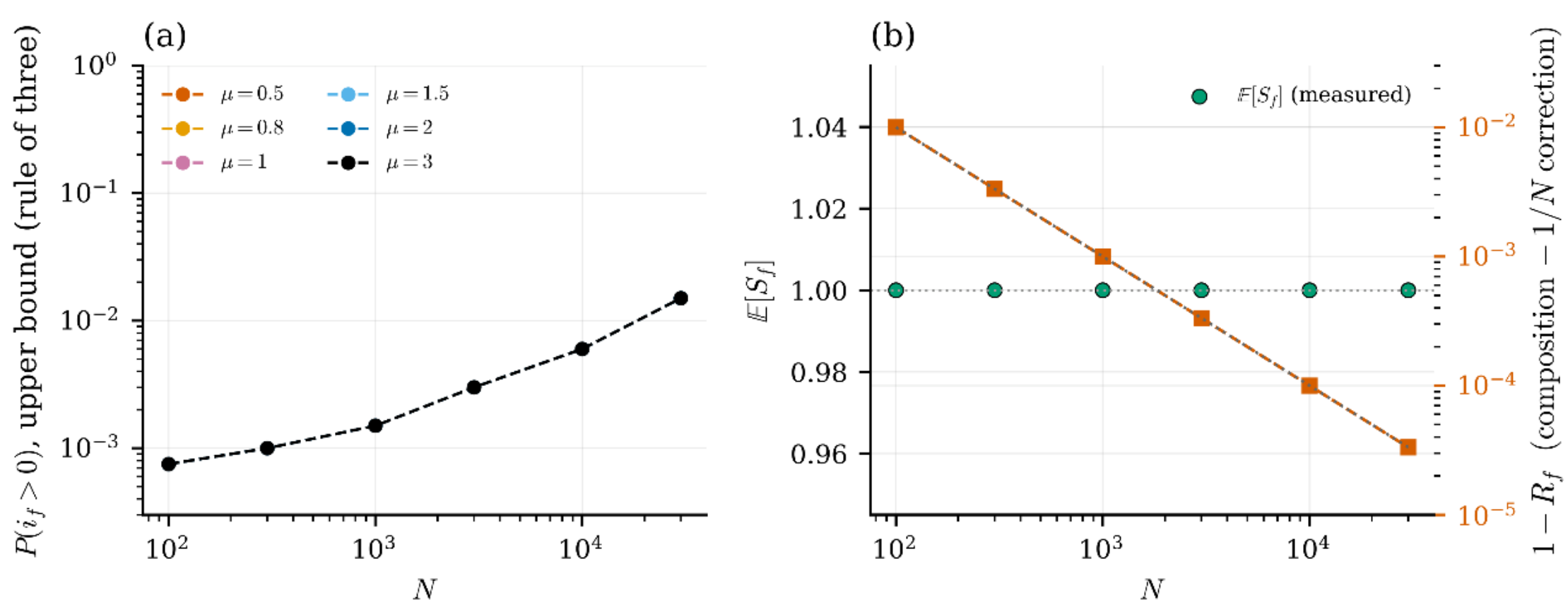

*Figure 8: The lone spreader. (a) Measured* $P(i_f > 0) = 0$ *at all* 36 $(N, \mu)$ *configurations; plotted symbols are the* 95% *upper bounds (rule of three,* $3/M$*). (b) Terminal state*

*$(i_f, n_f) = (0,1)$: the mean isolated-spreaders count $E[n_f]$ (green, left axis) sits exactly at $1$; the resolved composition correction $1 - R_f$ (orange, right axis) follows $1/N$ (dotted).*

## 7. Testable predictions

### 7.1 Archival inversion of the re-activation rate

The final-size law turns a cascade archive into a kinetic measurement. For a campaign (or a platform) in which single-burst and multi-burst accounts can be distinguished (precisely the $x_1/x_2$ definitions), the estimator

$$\hat{\alpha} = \mu \frac{\hat{x}_2}{\hat{x}_1} = \mu \frac{\text{re-activated silenced accounts}}{\text{single-burst silenced accounts}} \qquad (10)$$

recovers the re-activation rate from terminal data alone, with no time-series fitting. The error budget combines three terms measured in this paper: Monte Carlo noise of the final-size law ($\leq 0.95\%$ at $N = 6 \times 10^4$); the finite-population bias $-1/N$; and the network bias, which at the resolved composition level scales as $\langle k \rangle^{-0.71}$ (12% at $\langle k \rangle = 4$, 8.5% at $\langle k \rangle = 8$, below 3% at $\langle k \rangle = 40$; Figs. 6 and 7). Reporting an inferred $\alpha$ from real cascades therefore requires reporting an effective degree; the inversion is a *calibrated* measurement, and the calibration is the failure boundary itself.

### 7.2 Exploratory feasibility check on real cascades: Higgs-Twitter

We ran the inversion pipeline on public retweet-cascade data: the Higgs-Twitter activity records of De Domenico et al. (SNAP) [26], which timestamp 354,930 retweet events among 456,626 users over one week (a scientific "rumor" in the literal sense). Mapping a cascade to the retweet events directed at one source account, a participant's *spreading history* to the sequence of their retweet episodes (gap threshold $\theta$), and retirement to their last event, the terminal composition is measurable directly and the trajectory law can be tested a posteriori ($\Delta S/\Delta i = a - \hat{\mu}\,(S/i)$ with spreading-history intervals extended by a dwell window $\tau = \theta$).

Two assumptions and one robustness probe carry the model-to-archive mapping. (R1) *Silence = retirement*: an account is retired from the cascade from its last retweet episode onward; right-censoring at the one-week window edge is handled by dropping cascades whose last event lies within 24 h of the data edge. (R2) *Re-burst = re-activation*: episodes separated by more than $\theta$ are re-activations, so $S_1/S_2$ spreading histories are read as single-/multi-burst accounts. The mapping is one-way: it reads re-burst episodes as re-activation but cannot certify that a re-burst was contact-driven (the mechanism null, Sec. 7.4); a re-burst mechanism acting through extra emissions rather than re-exposure re-activation would enter the estimator identically, so $\hat{\alpha}$ is an effective rate that upper-bounds contact-coupled re-activation. Histories with three or more episodes enter $x_2$ once (the second episode creates the re-activation transition; later episodes change nothing). (R3) Threshold sensitivity: the $\alpha$-small conclusion is robust across the sensitivity grid $\theta \in \{2,6,24\}$ h, but the point estimate is not threshold-invariant: the cascade-aggregate inversion gives $\hat{\alpha}$ = 3.12% [2.50, 3.84]% ($\theta$ = 2 h), 2.21% [1.63, 2.95]% ($\theta$ = 6 h), and 0.39% [0.21, 0.61]% ($\theta$ = 24 h), an ≈8× spread tracking the truncation of long bursts by $\theta$; all calibrated statements below are anchored to the matched threshold $\theta$ = 6 h. $\hat{\hat{\mu}}$ varies even more ($1.60 \rightarrow 1.37 \rightarrow 0.32$ across the grid, crossing the resonant line $\mu = 1$), so R3 covers the re-activation conclusion only. Of 1388 sources with ≥ 30 episodes, the span guard ($\text{span} \geq 2\theta$) removes $19/86/605$ ($\theta = 2/6/24$ h), the dormancy guards (a cascade is admitted only if its last episode precedes the end of the observation window by ≥ 24 h, so that dormancy is demonstrated before the terminal composition is read) a further $497/493/448$, and the regression gate accepts $328/298/126$ cascades (37% of guard survivors at $\theta = 6$ h). The terminal composition at $\theta = 6$ h counts 49,738 single-burst and 800 multi-burst stifled accounts ($\hat{x}_2 = 1.58\%$ cascade-aggregate over the 50,538-account ledger, one per cascade; per-cascade lone-spreader censoring is applied at the regression stage). Three results (Fig. 9):

(i) *The identity check is diagnostic.* Across 298 cascades (≥ 30 participants, $\theta = 6$ h), the fitted intercept clusters at $a = 1.08$ (median; theory value 1, parameter-free), with per-cascade $R^2$ median 0.34 (interquartile range [0.14,0.55]); at $\theta = 24$ h the median is 1.04. At $\theta = 2$ h the check fails ($a = 0.41$): the dwell window is then too short to reconstruct activity, so the check itself diagnoses when the reconstruction is valid. The check validates the resolved composition construction in order of magnitude and in its $\theta$ trend, but not exactly: at matched parameters ($\mu = 1.37, \alpha = 0$) the synthetic archives of Sec. 7.3 give intercepts 2–3 × lower (0.14 at $\theta = 2$ h, 0.48 at $\theta = 6$ h), a systematic residual attributable to within-session clustering and dwell-window reconstruction that the time-warped synthetic archives do not reproduce.

(ii) *The resolved composition is dominated by single-burst histories.* The cascade-aggregate re-activated share is $\hat{x}_2 = 1.92\%,\ 1.58\%,\ 1.18\%$ at $\theta = 2{,}6{,}24$ h (cluster-bootstrap 95% CI [1.57, 2.31]%, [1.20, 2.09]%, [0.74, 1.71]%; the per-cascade median share is 0% with bootstrap upper end 0.5% at $\theta = 6$ h), and the trajectory regression gives $\mu = 1.37\ \ [1.27{,}1.53]$. The cascade-aggregate inversion (10) places the effective re-activation rate at $\alpha = 2.21\%$ ([1.63%, 2.95%]; joint cascade-level cluster bootstrap), the cascade-aggregate estimate is dominated by the minority of multi-burst histories, and the per-cascade distribution is zero-inflated (median and lower quartile 0, upper quartile 2.2% at θ=6 h): the per-cascade $\hat{\mu}$ dispersion (interquartile range 0.62–2.07 at $\theta = 6$ h) is the dominant uncertainty and is covered by the cluster bootstrap. Within assumptions R1–R3 the recruitment rate exceeds the effective re-activation rate by $\approx 45\times$, about one and a half orders of magnitude, not two; the calibration correction of Sec. 7.3 lowers it to $\approx 22 - 30\times$. The regression gate (≥30 participants) retains 298 of 809 candidate cascades (37%); it is size-based rather than α-based, and the C3 control (Sec. 7.4) shows the α-small conclusion is stable under activity-preserving shuffling.

(iii) *Conversations re-engage more than rebroadcasts.* On the reply-cascade view of the same records (8 cascades qualify), $\hat{x}_2 \simeq 3.2\%$ and $\alpha \simeq 4.3\%\ \ (per-cascade\ median;\ bootstrap\ CI\ [0\%, 10.9\%];\ one\ cascade\ reaches\ 22.4\%)$; the reply view re-engages more at the per-cascade caliber (cascade-aggregate caliber: reply $\hat{x}_2$ 2.5% vs. retweet 1.6%; reply $\hat{\alpha}$ 2.5% $[1.1\%, 10.0\%]$ vs. retweet 2.2% $[1.6\%, 3.0\%]$), consistent with conversation threads sustaining re-engagement, though with $n = 8$ the two views are compatible within errors and the comparison stays exploratory. Censoring (one-week window), source-level (not tweet-level) aggregation, and the network dressing of $\hat{\mu}$ (Sec. 5.2) are the dominant systematics; the feasibility check establishes that the final-size pipeline, with its built-in identity check, runs end-to-end on real archives. What the pipeline *measures* under these assumptions is calibrated next, on simulated archives with known parameters (Sec. 7.3) and by negative controls (Sec. 7.4).

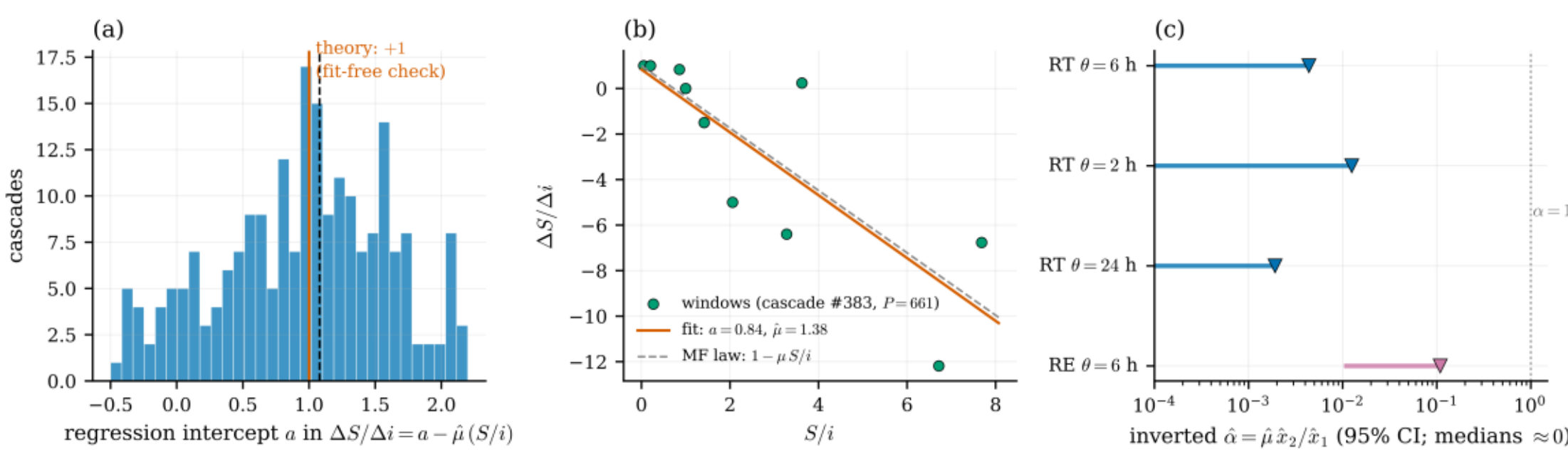

*Figure 9: Real-cascade demonstration (Higgs-Twitter, retweet view). (a) The parameter-free identity check: regression intercepts vs. the theoretical value* $1$ *(median* $1.08$, $N = 298$ *cascades,* $\theta = 6$ *h). (b) The trajectory law* $\Delta S/\Delta i = 1 - \hat{\mu}(S/i)$ *on a representative cascade (*$P = 661$, $R^2 = 0.67$*): the fit (solid) tracks the mean-field line (dashed). (c) Inverted re-activation rate:* $95\%$ *CIs (medians* $\approx 0$*) across session thresholds (pooled 3.12%, 2.21%, 0.39% at* $\theta = 2, 6, 24$ *h).*

### 7.3 Simulated archives with known parameters: end-to-end validation

The measurement model above maps model events onto platform episodes, and the mapping itself rests on assumptions that deserve a direct test. We therefore generated platform-style archives with known parameters and ran the same pipeline unchanged. The individual-level simulation uses event rates identical to the engine of Appendix D (recruitment $k_1 iS/N$; re-activation $\alpha k_1 s_1(S-1)/N$; stifling $\mu k_1 s_{1,2}(S-1)/N$; counts $n_I, n_{S_1}, n_{S_1,S_2}$ in place of population fractions), with $N$ log-uniform in [120,2000] and one archive generator per cell (a 10-generator-seed robustness sweep is reported below); recruitment and re-activation are recorded as retweet-episode events, while silence leaves no trace. A monotone rescaling of event times, $t \mapsto T^{1-\beta} t^{\beta}$ ($\beta = 0.5$), reproduces the bursty inter-event timing of human platform activity; the $(i, S)$ orbit is unchanged, and only the regression binning and the session-gap structure are affected. The full pipeline (session threshold $\theta$, a posteriori retirement, lone-spreader censoring, linear-bin trajectory regression, and inversion) was then run exactly as on real data (Fig. 10). This archive scale validates the pipeline's mechanics, not its asymptotics; the calibration is therefore used as an order-of-magnitude correction only (Sec. 7.2), and larger synthetic archives are the natural extension.

Three results stand out. First, the calibration produces no false positives: for $\alpha = 0$ the inversion returns $\hat{\alpha} = 0$ at the median in all nine $(\mu, \theta)$ cells ($\mu \in \{0.5, 1.37, 2\}$, $\theta \in \{0.5, 2, 6\}$ in $k_1 = 1$ units). Second, recovery is monotone: $\hat{\alpha}$ increases with the true $\alpha \in \{0.005, 0.01, 0.02, 0.04, 0.08\}$ in every cell (all 15 ($\mu$, $\alpha$) branches at each of the three $\theta$ values), while the resolution is limited: the interquartile ranges of adjacent $\alpha$ values overlap for $\theta \leq 2$ and $\mu \geq 1.37$ at $\alpha \leq 0.02$ (e.g. at $\mu = 1.37$ , $\theta = 2$ : IQR $[0.18\%, 1.02\%]$ at $\alpha = 0.01$ vs. $[0.52\%, 2.33\%]$ at $\alpha = 0.02$); the levels are resolvable for $\mu = 0.5$ throughout, and for $\alpha \geq 0.04$. The session threshold is anchored, not exactly matched: the synthetic thresholds $\theta \in \{0.5, 2, 6\}$ are expressed in simulation time units ($k_1 = 1$) and are matched to the real thresholds {2, 6, 24} h by residence-window role (short, medium, long), rather than by unit conversion; by equal cascade-aggregate re-activated share the closest calibration point is $\theta \approx 0.2$ (1.46% vs. the real 1.58% at $\theta$=6 h), while the grid anchor $\theta$=2 gives 1.0–1.2% and $\theta$=0.5 gives 1.34%; the share residual is carried as part of the calibration uncertainty, and the bias bounds are threshold-dependent (at $\theta = 6$ the $\mu = 1.37$ branch underestimates $\alpha$ by 53–100%), so Sec. 7.2 propagates the $\theta$=2 calibration — median relative bias −40% across the 15 branches (range −73% to +55%; the $\mu$=1.37 branches relevant to the real $\hat{\mu}$=1.37 underestimate by −33% to −52%) — as a correction of ×1.67 at the grid median (×1.5–2.1 on the $\mu$=1.37 branches) to the real-data estimate: an order-of-magnitude bound rather than a point estimate, since per-branch factors range from ×0.65 (overestimating $\mu$=0.5 branches) to ×3.7. Third, the calibration bias is bounded: at the matched threshold $\theta = 2$ the median relative bias of $\hat{\alpha}$ lies within $[-73\%, +55\%]$ across the grid (upward for $\mu = 0.5$, downward for $\mu \geq 1.37$).

We attribute it to the merging of re-activation episodes with the first burst in the session structure and to selection by the regression gate (which passes 16%–86% of synthetic cascades, with the same linear bins as for real data); at θ=2 the interquartile ranges of adjacent levels overlap for all 12 pairs across the three μ, and the level medians are strictly ordered everywhere. Degraded archiving leaves the estimates stable: truncating each cascade to 90% or 75% of its span, or dropping 15% or 30% of episodes, shifts the median $\hat{\alpha}$ by at most 0.11 percentage points (≤7% relative) from the uncensored baseline (Fig. 10(c)). Seed robustness. Repeating the matched-threshold grid (θ=2) with 10 independent archive-generator seeds (60 archives per branch and seed; results/w7_seed_variation.json in the code release) leaves the α=0 branches at exactly zero and the μ=0.5 and μ=1.37 branches monotone in every seed; at μ=2, five of ten seeds show adjacent-level inversions below α=0.04 — the resolution-limited corner already flagged by the overlapping IQRs. Branch medians vary across seeds by up to a factor ≈2 at the weakest branch (μ=1.37, α=0.02: per-seed medians in [0.76%, 1.62%]), within the calibration-bias band reported above.

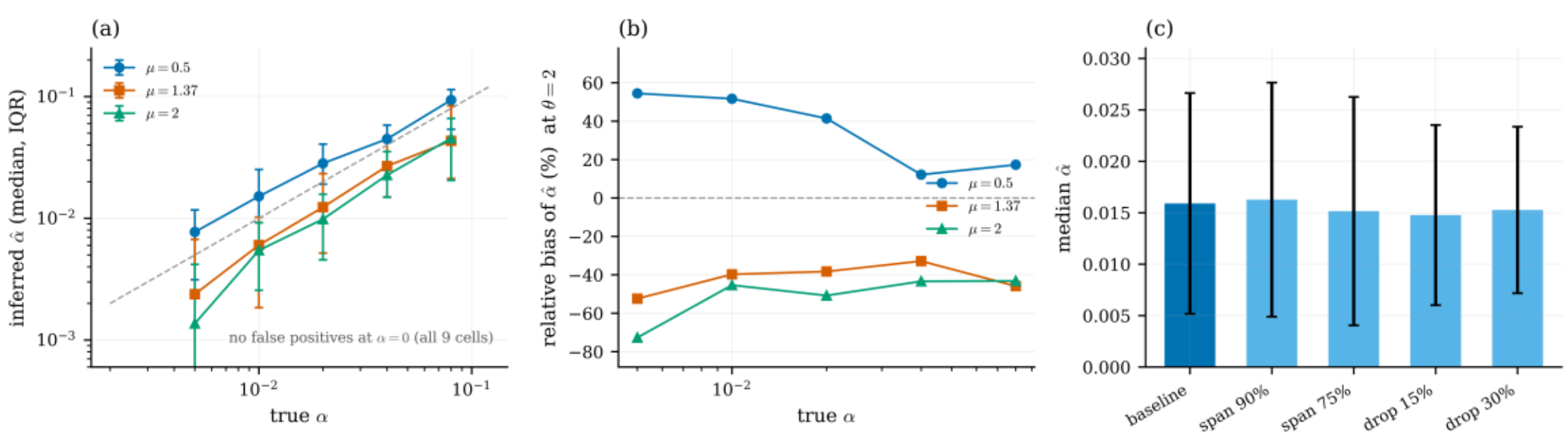

*Figure 10: Simulated-archive validation of the inversion pipeline (individual-level simulations with known parameters; event rates as in Appendix D; N log-uniform in* $[120,2000]$*; monotone rescaling of event times,* $t \mapsto t^{0.5}$*, mimicking the bursty timing of platform activity). (a) End-to-end recovery at the matched session threshold* $\theta = 2$ *(*$k_1 = 1$ *time units): inferred versus true re-activation rate, median with interquartile range over cascades passing the regression gate, for* $\mu \in \{0.5, 1.37, 2\}$*; the dashed line marks perfect recovery, and all* $\alpha = 0$ *cells (points on the axis) return* $\hat{\alpha} = 0$*, i.e. no false positives in any of the nine* $(\mu, \theta)$ *cells (error bars in (a): per-cascade interquartile ranges, not bootstrap CIs). (b) Median relative calibration bias versus true* $\alpha$ *at* $\theta = 2$*:*

*upward for* $\mu = 0.5$*, downward for* $\mu \geq 1.37$*; session merging drives the stronger downward bias at the long threshold* $\theta = 6$*. (c) Archival degradation (*$\mu = 1.37$*,* $\alpha = 0.02$*,* $\theta = 1$*): truncating each cascade to* $90\%$ *or* $75\%$ *of its span, or independently dropping* $15\%$ *or* $30\%$ *of episodes, shifts the median* $\hat{\alpha}$ *by less than* $7\%$ *(error bars: interquartile range).*

### 7.4 Negative controls

Four controls test whether the pipeline measures re-activation or an artifact of the archival preprocessing (retweet view, θ = 6 h, same guards). (C1) Timestamp shuffling within each cascade still passes the identity check and leaves $\hat{\mu}$ unchanged (1.33 vs. 1.37), but raises the cascade-aggregate re-activated share from 1.58% to 2.23%: scattering episodes in time creates multi-burst histories, so the estimator responds to temporal structure, not just burst counts. (C2) Source shuffling destroys cascade coherence: the identity check degrades (median $R^2$ = 0.24) and the re-activated share collapses to 0.01%. The inversion tracks cascade-level structure and fails without it. (C3) Activity-preserving shuffling leaves the α-small conclusion intact ($\hat{\alpha}$ = 0.25%): the estimator is not an activity-heterogeneity artifact. (C4) Burst-label randomization leaves the cascade-aggregate totals, and hence $\hat{\alpha}$, unchanged by construction; the informative nulls are the temporal and structural ones.

On the synthetic side, the $\alpha = 0$ archives of Sec. 7.3 provide the exact no-re-activation null (zero false positives at all nine cells). A *mechanism null* separates contact-coupled re-activation from uncoupled re-bursting: injecting independent re-burst episodes (rate $\lambda = 5 \times 10^{-4}$ per retired account per unit warped time, no contact coupling) into $\alpha = 0$ cascades raises $\hat{\alpha}$ to a comparable level ($0.025$ vs. $0.019$ median for contact-coupled re-activation at matched injection rate, not matched re-activation shares: the two populations differ in baseline re-burst timing (median activity level at re-burst $0.08$ vs. $0.52$)), confirming that terminal compositions alone cannot distinguish the two mechanisms. The re-burst *timing* does discriminate them: the Pearson correlation between the re-burst timing density and the pool's activity profile $S(t)/S_{\max}$ is $\rho = +0.37$ (median) for contact-coupled re-activation ($\alpha = 0.02$) versus $\rho = -0.13$ for uncoupled re-bursting, with non-overlapping interquartile ranges ($+0.22$ to $+0.59$ vs. $-0.25$ to $-0.07$; the full ranges do overlap). Re-activation follows the pool's activity profile, whereas independent re-bursting does not. Terminal compositions identify the effective re-activation rate; identifying its mechanism requires timing information beyond the terminal state.

## 8. Discussion

**Validity domain.** The three laws are exact for the mean field of the locked two-stage model: irreversible channels only, homogeneous mixing (or class-wise HMF), static or label-blind graphs. The S×R analysis of Sec. 4.3 fixes what breaks first when the classic stifler-contact channel is restored, and the network study fixes how much breaks with structure. The $-1/N$ correction of Theorem 8 ($n_0 = \varepsilon N$) is the entire finite-population correction for the driver.

**Multi-stage extension to $f \geq 3$ stages.** The common-factor cancellation is indifferent to the number of stages: an $f$-stage chain produces $f - 1$ history-resolved layers, each a linear equation driven by its own rate sum $\gamma_j = \mu + \alpha_j$, and the terminal composition obeys the nested-product conjecture $x_m = \prod_{j<m} \frac{\alpha_j}{\mu+\alpha_j} \cdot \frac{\mu}{\mu+\alpha_m}$ for $m \leq f$ (with $\alpha_f \equiv 0$), verified numerically to $10^{-10}$ for $f = 3$ across six $(\mu, \alpha_1, \alpha_2)$ configurations (results/w8_f3_conjecture.json in the code release); $f = 2$ reduces exactly to Theorem 2, non-identifiability extends unchanged (the merged layer remains $dS/di = -1 + \mu S/i$), and the inversion becomes sequential. The closed forms generalize; the present two-stage case is the smallest non-trivial member, retained here because it is the level the data resolve (the archival distinction single-burst/multi-burst is exactly two-stage).

**Mechanism heterogeneity.** The model assigns one re-activation rate $\alpha$ to all re-exposures; real platforms mix fatigue, attention cycles, bot activity, algorithmic promotion, and synchronized external events. The theorems should be read with $\alpha$ as an *effective* re-activation rate under the measurement model of Sec. 7.2: the final-size law and its linear inversion survive any heterogeneity that re-activates accounts at a rate proportional to the pooled spreading activity, while mechanism-level distinctions (Sec. 7.4, the mechanism null) require timing information beyond the terminal composition. Individual activity multipliers (agent-specific rate rescaling) break the exact cancellation unless they act equally on all channels of an event; in the annealed limit the pooled laws survive with the activity-weighted pool replacing $S$, and the estimator inherits $\langle k^2 \rangle$-type corrections analogous to Sec. 5.2. Which structural statements (non-identifiability, saturation) survive monotone label-preserving heterogeneity is an open problem; the $\langle k \rangle^{-0.71}$ network boundary and the $g$ † gates are the first quantitative map of the exact domain.

**Transferability.** The transferable object is a mechanism, not an analogy: whenever every channel of an A2SP process carries the same partner-pool factor, (a) trajectory equations close without the common factor, (b) terminal inventories freeze at ratios of rate sums, and (c) parameters acting only on internal transitions vanish from all marginal statistics. Step-growth polymerization realizes the mechanism with a consumable common factor ($[A]$); rumor spreading, with a catalytic one ($S$). The same skeleton is worth scanning for in other social processes built on A2SP events [23].

## 9. Conclusion

This paper has imported the common-factor-cancellation mechanism of the A2SP process into a two-stage Maki–Thompson rumor model and derived three exact terminal-state theorems. (i) Saturation: for every stifling-to-recruitment ratio $\mu > 0$ the rumor exhausts the ignorant pool, with a three-way tail classification and a logarithmically critical line at $\mu = 1$. (ii) Final size: the terminal stifler pool splits by spreading history as $(x_1, x_2) = [\mu/(\alpha+\mu), \alpha/(\alpha+\mu)]$, independent of seed size, and the fingerprint inverts linearly, $\alpha = \mu x_2/x_1$. (iii) Pooled versus history-resolved: pooled statistics are exactly independent of α on any static graph, while only the history-resolved composition carries α. Each theorem carries a quantified boundary of validity: the classic teller-retires channel breaks the exact closure beyond quantified gates; on configuration-model networks the composition bias decays as a ⟨k⟩^−0.71 power law; and the finite-population correction is a single isolated spreader frozen in the absorbing state. Empirically, a parameter-free identity check and four negative controls support the end-to-end inversion of 354,930 Higgs-Twitter retweet episodes, yielding a model-conditional estimate $\hat{\alpha} = 2.21\%$ (≈3.7% after calibration). Reach-type interventions are provably α-free and need no re-activation data, whereas history-resolved objectives require the terminal composition; the two must not be mixed in a single objective function. Terminal compositions can measure the effective re-activation rate; identifying its mechanism requires timing information beyond the composition.

**Acknowledgements**

During the preparation of this manuscript, the authors utilized GLM-5.3 flash, developed by Z.ai, and Doubao, as a computational and writing aid. Specifically, the AI tool was used to assist with algebraic simplifications and validations during the theoretical derivations and to improve the English language readability of the text. The authors independently derived, recalculated, and verified all kinetic equations, mathematical proofs, and physical boundary conditions. The authors take full responsibility for the theoretical rigor, scientific accuracy, and final content of this manuscript.

**Data availability**

The symbolic kernel symer (v0.1.3), including all regression anchors, the review-revision reproduction scripts, the Gillespie SSA numerical core, the inversion CLI for real cascade resolved compositions, and the figure-level source data (results/*.json), is open source under the MIT license at https://github.com/QINGQUAN-T/symer and permanently archived on Zenodo [28] (concept DOI: 10.5281/zenodo.23282847, which resolves to the latest release). Real-cascade experiments use the public SNAP Higgs-Twitter dataset [26]; every number reported in the paper is reproduced by the released scripts, whose figure-level source data and generation scripts are archived in results/*.json and scripts/. To reproduce: pip install -r requirements.txt; python run_tests.py (49/49 regression tests); the scripts/w7_*.py regenerate every figure-level JSON (original run logs in logs/). data_README.md documents the SNAP download and the rename to data/higgs-activity.txt.gz.

## CRediT authorship contribution statement

**Qingquan Tang:** Conceptualization, Methodology, Formal analysis, Software, Validation, Investigation, Data curation, Writing – original draft, Writing – review & editing, Visualization.

## Declaration of competing interest

The author declares that he has no known competing financial interests or personal relationships that could have appeared to influence the work reported in this paper.

Ethics statement. This study uses only fully anonymized, publicly released third-party data (SNAP Higgs-Twitter); no human-subjects data were collected by the author, and no additional ethical approval was required under the source release terms.

Funding. This research did not receive any specific grant from funding agencies in the public, commercial, or not-for-profit sectors.

## Appendix A. Proof of the tail trichotomy (Theorem 1)

Along trajectories: from (3), $S/i^{\mu} = C_S - i^{1-\mu}/(1-\mu)$ exactly, and $i = -k_1 i S$, so

$$\frac{d}{dt} i^{-\mu} = \mu k_1 \frac{S}{i^{\mu}} = \mu k_1 \left( C_S - \frac{i^{1-\mu}}{1-\mu} \right).$$

For $\mu < 1$ the right-hand side increases monotonically to $\mu k_1 C_S$; integrating, $i^{-\mu} = i_0^{-\mu} + \mu k_1 C_S\, t\big(1 + o(t)\big)$, i.e., $i \sim (\mu k_1 C_S t)^{-1/\mu}$, and $S = C_S i^{\mu}\big(1 + o(1)\big) = (\mu k_1 t)^{-1}\big(1 + o(1)\big)$. For $\mu > 1$ the linear term dominates, $S \approx i/(\mu - 1)$, giving $d(1/i)/dt = k_1 S/i \to k_1/(\mu - 1)$ and $i \sim (\mu - 1)/(k_1 t)$. For $\mu = 1$, $u = 1/i$ obeys the exact equation $du/dt = k_1(C_S + \ln u)$, whose logarithmic-integral asymptotics give $u \approx k_1 t[\ln(k_1 t) + \ln\ln(k_1 t) + C_S - 1]$, i.e., $i \sim 1/(k_1 t \cdot lnt)$. In all three cases $S \to 0$ and $i \to 0$ as $t \to \infty$ with $i(t) > 0$ on finite intervals; together with the monotonicity of $i$ this proves Theorem 1. The pooled layer identity $x_1 + x_2 = 1$ used in Theorem 2 follows from the same bookkeeping applied to $\int_0^{i_0} (S/i)\ di = (i_0 + \varepsilon)/\mu$.

## Appendix B. Proof of architectural non-identifiability (Theorem 6)

Order the state space by the pooled variables $q = (i, S_{count}, r)$ together with the graph and the edge-type counts. Transitions come in three families: recruitment ($I \to S_1$ on some $S$–$I$ edge), stifling ($S_k \to R$ on some ordered spreader pair), and re-activation ($S_1 \to S_2$ on some ordered $S \to S_1$ pair). Their total rates depend on the state only through the numbers of edges of each type (e.g., $k_1 E_{SI}$, $k_2 \cdot E_{SS}^{ord}$), where E_SS^ord counts ordered spreader–spreader pairs, so k₂·E_SS^ord = k₂·S(S−1) tends to the mean-field k₂S² as N→∞ (the ordered-pair convention absorbs the self-exclusion); none of which involves $\alpha$ or the $S_1/S_2$ split beyond the pool; and a re-activation event changes labels only, leaving $q$ and every edge-type count invariant. The generator therefore commutes with all label permutations within the pool, and the lumped process on $q$ is Markov with an $\alpha$-free generator; standard lumpability theory then gives equality in distribution of the pooled process for all $\alpha$. The same argument applies verbatim to a time-dependent graph whose evolution is label-independent, and class-wise to the heterogeneous mean-field system, where $\dot{S}_k = k k_1 i_k \Theta_S - k k_2 S_k \Theta_S$ is $\alpha$-free.

## Appendix C. S×R perturbation: numerics

With the channel $S \times R \to R + R$ at rate $g k_1 s_k r$ (teller retires), the exact merged solution (9) was verified by computer algebra (symbolic residual 0; $g \to 0$ recovers (3)) and its crossing root $i_f(g)$ matches the numerically integrated partial saturation to six decimals for $\mu \in \{0.5, 2\}$, $g \in [0.05, 0.5]$. Gillespie simulations ($N = 6 \times 10^4$, six seeds) agree with the mean-field $i_f(g)$ within 1.5%. Composition deviations from (7) (the resolved composition fluxes are accumulated as extra ODE states, in exact agreement with the Gillespie counts to three decimals) cross the ±2% acceptance gate at $g \dagger (\mu)$: spreading history $x_1$ at 0.0147 ($\mu = 0.5$), 0.0171 ($\mu = 1$), 0.3174 ($\mu = 2$); stifler share at 0.0291 ($\mu = 0.5$), 0.0147 ($\mu = 1$), 0.0083 ($\mu = 2$). The gates are non-monotone in $\mu$: the spreading history $x_1$ channel is most robust near $\mu = 2$, where the $x_1$ law value $\mu/(\alpha + \mu) = 2/3$ sits farthest from the perturbation-induced drift. The merged-layer $\alpha$-non-identifiability holds to $\leq 3 \times 10^{-12}$ at all $g \in \{0, 0.1, 0.5\}$. The gate values are collected

in Table 1 in Appendix C (in full: μ=0.5: 0.0147/0.0291; μ=0.75: 0.0152/0.0195; μ=1: 0.0171/0.0147; μ=1.5: 0.0289/0.0102; μ=2: 0.3174/0.0083; μ=3: 0.1065/0.0067, resolved $x_1$ / stifler share). For $g > \mu$ the closed form (9) is used only on $(i_f(g), i_0)$: the frozen residual $i_f(g)$ is positive ($i_f = 0.075$ at $\mu = 0.5$, $g = 0.6$; $0.257$ at $g = 2$) and the closed form stays non-negative on the whole bracket, so the physical branch never samples the divergent sector $C\, i^{\mu-g}$ near $i = 0$.

*Perturbation gate $g\dagger$ of the classic $S \times R$ channel: the value at which the terminal composition departs from the final-size law (7) by more than $\pm 2\%$, computed by exact mean-field integration with ODE-accumulated resolved composition fluxes (this appendix).*

| μ | g† gate for career $x_1$ | g† gate for stifler share |
|---|---|---|
| *0.5* | 0.0147 | 0.0291 |
| *0.75* | 0.0152 | 0.0195 |
| *1.0* | 0.0171 | 0.0147 |
| *1.5* | 0.0289 | 0.0102 |
| *2.0* | 0.3174 | 0.0083 |
| *3.0* | 0.1065 | 0.0067 |

**Appendix D. Numerics and reproducibility**

Mean-field integrations use LSODA ($rtol = 10^{-11}$, $atol = 10^{-15}$), tail windows $t \leq 2 \times 10^5$, and verification criteria that require no fitted parameters (exact identities, transform linearizations, plateau convergence). The network engine is an event-driven CTMC with three event buckets (swap-delete arrays with position dictionaries: $O(1)$ sampling, $O(\deg)$ incremental bookkeeping); a stifling event on an $S_1$–$S_1$ edge generates four event terms (both retirements and both re-activations), $S_1$–$S_2$ three, $S_2$–$S_2$ two. Graphs are configuration models (self-loops removed, multi-edges collapsed, isolates deleted and relabeled; actual $\langle k \rangle$ reported). The HMF comparison uses the same degree distribution (theoretical Poisson pmf; empirical histogram for scale-free graphs). The well-mixed engine is the direct method with pair self-correction [$(S_{count} - 1)$ factors]. All random seeds, parameter grids, raw logs and structured outputs are fixed and archived with the code release; the seed protocols differ by figure (Fig. 6: 8 SSA seeds per point; Fig. 7: 10 graph realizations × 3 seeds; Figs. 3–5, 8–10: seeds recorded per output file in results/), and figure-level source data are archived as results/*.json; every number in this paper regenerates from the released scripts (v0.1.3).

## Appendix E.

**Notation.** Symbols used throughout; rates are per contact per unit time, fractions sum to one, and the two career classes partition the spreader pool.

| | |
|---|---|
| $i, S, r$ | ignorant, spreader, retired population fractions; $i + S + r = 1$ |
| $s_1, s_2$ | career fractions of the spreader pool ($S = s_1 + s_2$ on the career layer) |
| $x_1, x_2$ | terminal career ledger (frozen stifler-pool composition), $x_1 + x_2 = 1$ |
| $k_1, k_2$ | per-contact recruitment and stifling rates |
| $\mu = k_2/k_1$ | stifling-to-recruitment ratio (merged-layer knob) |
| $\alpha$ | arming (re-activation) rate, in units of $k_1$ |
| $\gamma = \alpha + \mu$ | career-layer rate sum |
| $\varepsilon$ | initial seed fraction ($\varepsilon \cdot N$ spreaders at $t = 0$) |
| $N$ | population size (well-mixed) or graph size (Sec. 5) |
| $g, g^\dagger(\mu)$ | restored classic S×R teller-retires channel strength, in units of $k_1$ (Sec. 4.3); $g^\dagger(\mu)$ marks the strength at which the ledger departs 2% from (7) (Table 1) |
| θ | episode gap threshold for session reconstruction (Sec. 7.1) |
| E_SI, E_SS^ord | counts of I–S edges and ordered S–S spreader pairs (Appendix B) |